\documentclass{article}

\usepackage{PRIMEarxiv}

\usepackage[utf8]{inputenc} %
\usepackage[T1]{fontenc}    %
\usepackage{hyperref}       %
\usepackage{url}            %
\usepackage{booktabs}       %
\usepackage{amsfonts}       %
\usepackage{amsmath}
\usepackage{nicefrac}       %
\usepackage{microtype}      %
\usepackage{lipsum}
\usepackage{fancyhdr}       %
\usepackage{graphicx}       %
\usepackage{subcaption}
\usepackage{caption}       %
\usepackage{svg}  %
\graphicspath{{figures/}}     %
\usepackage[table,xcdraw]{xcolor}
\usepackage{array}
\usepackage{booktabs}
\usepackage{siunitx}
\usepackage{xcolor} 
\usepackage{colortbl}
\usepackage{placeins}
\usepackage{soul,color}
\usepackage[backend=biber, style=nature]{biblatex}
\newcommand{\R}{\mathbb{R}}

\newcommand{\tens}[1]{%
  \mathbin{\mathop{\otimes}\limits_{#1}}%
}

\title{Facet: highly efficient $E(3)$-equivariant networks for accelerated training of interatomic potentials }

\author{%
  Nicholas Miklaucic$^1$, Lai Wei$^1$, Rongzhi Dong$^1$, Nihang Fu$^1$, Sadman Sadeed Omee$^1$, \\Qingyang Li$^1$, Sourin Dey$^1$, Victor Fung$^2$  Jianjun Hu$^{1}$\thanks{Corresponding author: jianjunh@cse.sc.edu}\\
  $^1$Department of Computer Science and Engineering, University of South Carolina, Columbia, SC, USA\\
  $^2$Computational Science and Engineering, Georgia Institute of Technology, Atlanta, GA, USA\\
  }
  
\begin{document}
\maketitle

\begin{abstract}
Computational materials discovery is severely constrained by the high expense of first-principles calculations. Machine learning (ML) potentials that accurately predict energies from crystal structures offer a promising solution, but current approaches face significant computational bottlenecks. Steerable graph neural networks (GNNs) encoding geometric information with spherical harmonics respect the fundamental symmetries of atomic systems—permutation, rotation, and translation—for more physically realistic predictions. Maintaining equivariance throughout these networks, however, presents substantial design challenges: standard network components such as activation functions require modification, and each network layer must accommodate multiple data types corresponding to different harmonic orders.
We introduce Facet, a novel GNN architecture for efficient ML potentials developed through systematic analysis of existing steerable GNN designs. Our key innovations include replacing compute-intensive multi-layer perceptrons (MLPs) for processing interatomic distances with efficient splines that achieve equivalent performance while dramatically reducing computational and memory requirements. We also propose a general-purpose equivariant layer that mixes node information via spherical grid projection followed by standard MLPs—an approach that outperforms iterative tensor products in speed and surpasses linear and gate layers in expressiveness.
Training on the MPTrj dataset, our model achieves performance comparable to leading approaches with significantly fewer parameters and less than 10\% of the training computation. We demonstrate a 2× speedup compared to MACE models on a crystal relaxation task representative of crystal structure prediction workflows. We validate our design through systematic experiments with existing steerable GNNs, demonstrating that SevenNet-0's parameter count can be reduced by over 25\% without performance degradation. These techniques enable more than 10-fold acceleration in training large-scale foundation models for ML potentials, potentially transforming the landscape of computational materials discovery.

\end{abstract}

\section*{Introduction}
First-principles-based atomic simulations such as Density Functional Theory (DFT) and ab initio Molecular Dynamics (MD) have been playing an increasingly important role in materials and chemical discovery \cite{louie2021discovering,oganov2019structure,neugebauer2013density}, providing critical tools to predict and understand material structures and properties at the atomic and electronic levels. Their wide application \cite{omee2024crystal,wang2024machine,omee2025polymorphism} has been severely limited, however, by their extremely high computational complexity, which can render them infeasible for simulating large complex crystals or molecules \cite{deringer2019machine,ramprasad2017machine}. 

To accelerate these atomic simulations and consequently the materials discovery process, machine learning interatomic potentials (MLIPs) have been developed that predict energies, forces, and other properties without the huge computational cost of first-principles calculations or the expense of experiments \cite{friederich2021machine,ruff2024cogn,liao2022equiformer,batzner2022nequip,batatia2024macemp0,park2024sevennet,belli2025efficient,roberts2024machine,wang2024enhancing}. Among these machine learning (ML) potentials, deep neural network models have demonstrated superior performance \cite{fedik2022extending,chen2022universal,wen2022deep,yang2025efficient} in modeling the complex underlying physics and predicting materials and chemical properties, providing a convenient mechanism to handle the symmetries of the problem via network design. An examination of the latest Matbench-Discovery leaderboard \cite{riebesell2023matbench} shows that deep neural network-based models have made tremendous progress in their performance for high-throughput materials discovery, specifically focusing on predicting the stability of inorganic crystals. Most performance metrics have been improved by 200\% in the past 30 months: for example, energy prediction error RMSD has been improved from 0.112 meV/atom for M3GNet \cite{chen2022universal} to 0.061 meV/atom for eSEN-30M-OAM \cite{fu2025learning}, MAEs reduced from 0.075 meV/atom to 0.018 meV/atom, with the combined performance score (CPS) increased from 0.428 (M3GNet) to 0.888 (eSEN). These advances are made possible through three main approaches of the foundational model strategy: data, model, and computing resources. The state-of-the-art model eSEN-30M-OAM is trained with 113M structures (from 66M stable structures) while M3GNet is trained with 118K structures (from 62.8K stable structures). For each performance metric, we found that the top four models were all trained with huge training sets containing more than 100M samples/structures, which also showed superior performance in downstream applications \cite{han2025benchmarking}. Training such models requires huge computational resources and time: for example, the MACE-MP-0 model (see Table 1) was trained on A100 for 310 days using 1.58M training samples while SevenNet-0 was trained for 90 days on A100. Such huge computing requirements make it challenging to quickly update models or iterate on models by exploring different training data, tuning model parameters, and training algorithms.
While the ML potential community has been focusing on improving the predictive capabilities, generality, transferability, and scalability of such ML potentials \cite{deringer2019machine,fedik2022extending,wang2024machine}, the importance of training large atomic foundation ML potential models makes it imperative to develop ML potential models that allow much more efficient training and inference, similar to the DeepSeek moment in large language models (LLMs) \cite{guo2025deepseek,wang2025review,deng2025exploring}. How can one train a high-performance model with less than 10\% of the computing resources required to train current SOTA foundation models? To address this issue, several recent works have been proposed that focus on parallelization and distributed computing architectures for ML potential acceleration \cite{fuchs2025chemtrain,han2025distmlip}. In this work, we focus on model architecture development for building efficient ML potentials.

One of the main strategies to improve model efficiency is increasing  sample and parameter efficiency by implementing transformation (rotation and translation) invariance or equivariance. For example, graph neural network (GNN) architectures have been widely used to ensure invariance to permutation of nodes and edges as well as rotation and translation. The periodic symmetries of crystalline materials can additionally be incorporated by using periodic graphs \cite{yan2024periodic}, with nodes representing a set of symmetrically equivalent atoms and edges representing relative spatial positions between atoms. The use of relative spatial positions instead of absolute positions reflects the underlying translational invariance: any origin is fundamentally arbitrary.

Many material properties also exhibit rotational symmetry, which is more challenging to ensure through neural network design. 
Early approaches relied on \textit{scalarization}: restricting the model inputs to invariants under rotation, in the way that the periodic graph construction ensures translation invariance by removing explicit coordinate information. CGCNN \cite{xie2017cgcnn}, for example, only considers the lengths of edges and discards directional information. This information loss necessarily limits the expressivity of the network, however, as shown through Weisfeiler-Leman isomorphism: there exist distinct graphs with different chemistry that element type and distance alone are insufficient to distinguish \cite{xu2019wliso}. Considering the importance of angular information for material stability, strongly incorporating angular information should provide useful inductive biases for a network.

To address these limitations, 
subsequent work \cite{gasteiger2020dimenetpp,gasteiger2021gemnet,choudhary2021alignn} has introduced increasingly sophisticated scalarization techniques: incorporating angles between triplets of nodes \cite{choudhary2021alignn} and even dihedral angles between quadruplets of nodes \cite{gasteiger2021gemnet}. These techniques are effective, but they result in increasingly complex and expensive architectures that scale poorly with the size of the system being considered.

Concurrently, a \textit{steerable} GNN approach \cite{batzner2022nequip,batatia2024macemp0,park2024sevennet,merchant2023gnome} has avoided scalarization entirely. Using irreducible representations (\textit{irreps}) of the group of 3D proper rotations $SO(3)$ based on spherical harmonics, the information of how an output varies under rotations to the inputs can be tracked. The Clebsch-Gordan tensor product combines these representations equivariantly, generalizing operations such as the dot product and cross product. By representing edges as spherical tensors, a simple graph convolutional network without complex scalarization schemes can still encode important geometric information. These models have proven successful in many problems with atomistic systems as input and include MACE-MP-0 \cite{batatia2024macemp0}, SevenNet \cite{park2024sevennet}, GNoME \cite{merchant2023gnome}, and NequIP\cite{batzner2022nequip}.

The simplicity of these network architectures, which do not need complex scalarization schemes or explicit many-order summations, masks the complexity of their underlying components. Most standard numerical operations do not preserve equivariance when applied to irreps. For example, a foundation of modern deep learning—the nonlinear activation function—has no direct equivalent for irreps, as applying element-wise operations to irreps depends on the arbitrary coordinate system. These limitations complicate the transfer of successful GNN architectures from other domains to materials property prediction. These constraints do simplify the development of new architectures: equivariance imposes strict restrictions on how information can flow and enables direct comparison of many existing MLIPs.

In this paper, we make several contributions to the design of steerable GNNs for significant acceleration of high-performance machine learning potentials:
\begin{itemize}
\item We identify that the most computationally expensive part of many steerable GNNs is the message filter used in equivariant convolution, which can be greatly simplified without significant performance impacts. We demonstrate this by modifying a pre-trained SevenNet model, SevenNet-0, reducing the parameter count by over 25\% with negligible change in accuracy.
\item We consider the trade-off between computational complexity and expressiveness in the part of the network that mixes information within an individual node. By applying multi-layer perceptrons (MLPs) to a regular spherical grid encoding directional information, we achieve a middle ground between the expensive tensor product operations used in MACE and the simpler nonlinear gate operations used in SevenNet and GNoMe.
\item Using the above insights, we propose and train a new deep learning potential named Facet, which achieves comparable high performance to much larger SOTA deep learning atomic potentials while using only a fraction of their parameter size and can be trained with less than 10\% of their training cost, e.g., reducing the training time from SevenNet-0's more than 90 days to 2 days when trained with the same number of training samples.
\end{itemize}

\section*{Results}

\subsection*{Model architecture overview of Facet}

We begin by delineating our Facet potential model from prior work. While we base our approach on the successful SevenNet \cite{park2024sevennet}, we differ from the previous state of the art in several ways. To update node representations between message-passing steps, we use a general, expressive $S^2$-MLP-Mixer layer explained below. This approach contrasts with the MACE model \cite{batatia2024macemp0}, which uses an expensive symmetric tensor product layer for the same purpose, and SevenNet/GNoMe/NequIP \cite{park2024sevennet,merchant2023gnome,batzner2022nequip}, which use nonlinear gate layers that do not mix non-scalars with each other. As in SevenNet, we apply the same node update to all nodes regardless of element type, which reduces the number of parameters required by two orders of magnitude and additionally mitigates overfitting.

All of the aforementioned deep learning potential models share a message-passing convolution with essentially the same architecture. We find that the MLP computing weights from edge distance, the most expensive part of message passing in most such networks—and by extension one of the most expensive parts of the network—is unnecessary and can be replaced with a linear layer without any significant change to performance.

\begin{figure}[h!]
 \includegraphics[width=\textwidth]{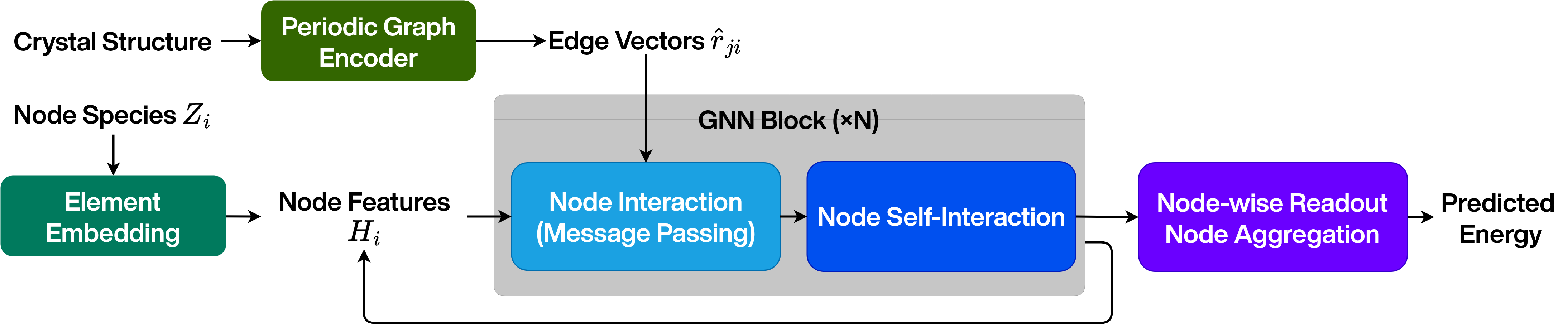}
\caption{Architecture overview of Facet. Information is stored only for individual nodes. Geometric information is incorporated through message-passing blocks, updating each node with information from its local environment. Information from that environment is then mixed within each node through a self-interaction block. After N=3 interactions and self-interactions, each node's embedding is a 128-dimensional vector processed to obtain energy.}
\label{fig:architecture-overview}
\end{figure}

These large changes, combined with many small optimizations designed with the aim of training efficiency without sacrificing expressivity, make our Facet architecture more effective at smaller sizes than previous networks trained on the same data, achieving both model parameter efficiency and training efficiency. The architecture of our model is shown in Figure \ref{fig:architecture-overview}. 

Facet encodes its input as a periodic orbit graph: each atom within the repeating unit cell is represented by a node. Edges are added to link each node with its 32 nearest neighbors: more than one edge can exist between two nodes, each referring to a different unit cell offset \cite{yan2024periodic}. Each node's atomic element is encoded with a 128-dimensional embedding trained from scratch. This initializes the node features as 128 scalars. Then, to incorporate geometric information, the interaction block combines this data with edge information. The interaction block updates the node features with incoming messages from neighbors to form node representations that contain non-scalar information in addition to scalars. Then this information is processed within each node by applying a node self-interaction layer to each node independently. The combination of message passing and self-interaction forms a single GNN block, which can be stacked to form a deeper network. For predicting energy or any other scalar, the last GNN block need only process scalar values, producing a 128-dimensional scalar embedding for each node. These are processed individually by a readout layer to predict node energies and then summed across a structure to obtain the predicted energy.

In the Facet model, we apply a normalization and a residual connection after each layer. Following EquiformerV2 \cite{liao2023equiformerv2}, we use equivariant layer normalization with scalars treated separately. This normalizes the magnitudes of each kind of irrep to average 1 and then scales by a learned irrep-wise constant. This is then added to a residual connection, padding the layer input with zeros if needed to match the shape of the layer output. By initializing the layer scale to 0, each GNN block becomes a simple identity function at the start of training. This initialization simplifies training for deep networks by ensuring block outputs start with approximately unit scale and zero mean.

To convert these approximately unit-scale outputs to energy, each node's energy prediction is scaled by a species-wise constant with a species-wise shift, added together across the crystal, and then scaled and shifted by a global value. These are set initially with simple linear regression but set as trainable to correct for any biases introduced by the simplicity of the original fit, which assumes average atom energy is independent of other atoms in the composition.

\subsection*{Interaction block: steerable graph convolution, streamlined}

\begin{figure}[h!]
\includegraphics[width=\textwidth]{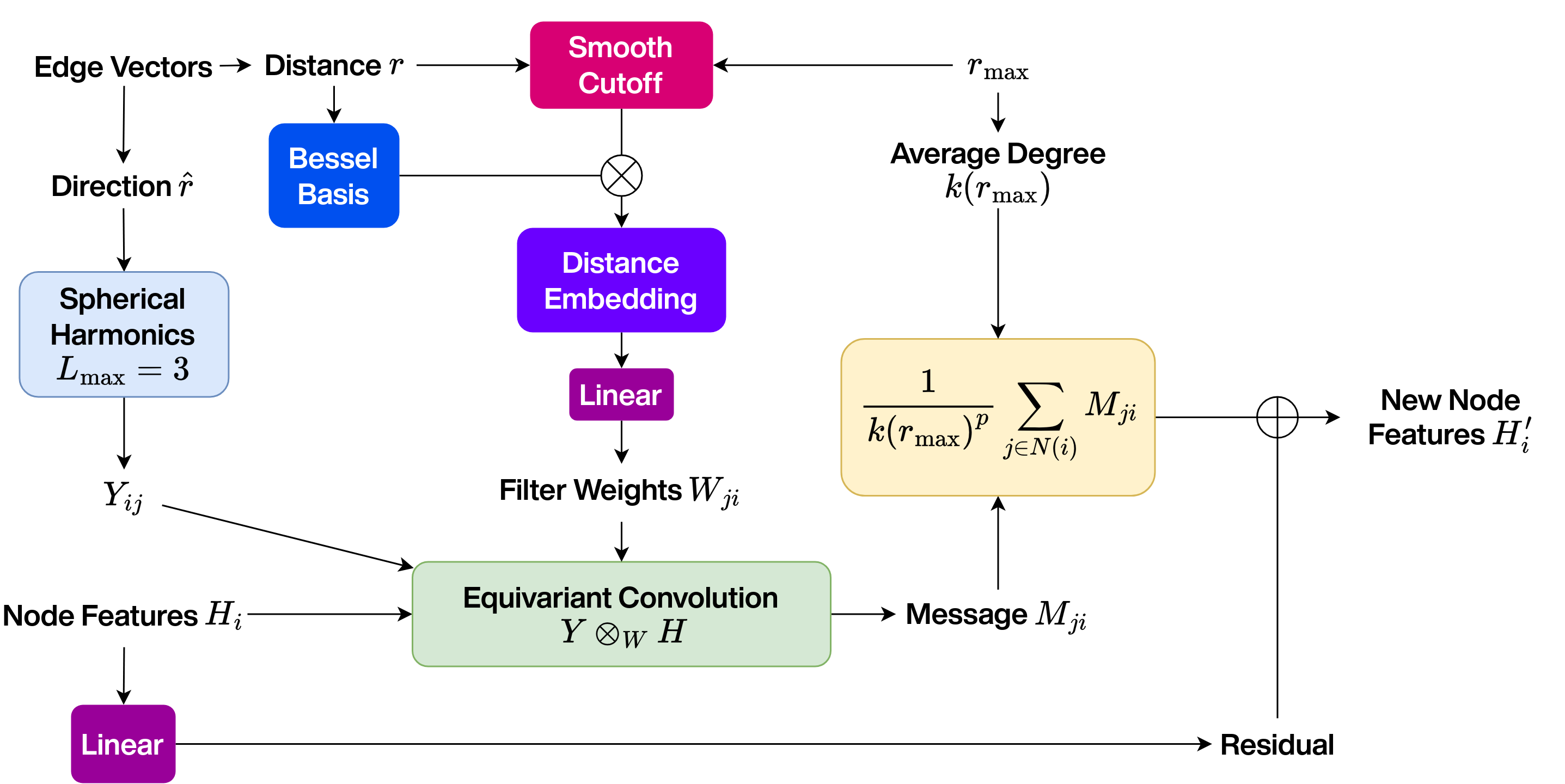}
\centering
\caption{The Facet interaction block. Edge vectors are decomposed into a directional component, embedded into spherical harmonics, and a distance component, encoded using a Bessel basis with a smooth distance cutoff so edges with $r \le r_{max}$ have no effect. The node features $H_i$ and the spherical harmonics $Y_{ij}$ are equivariantly combined using the Clebsch-Gordan tensor product. Each output is then weighted with a filter generated from a linear projection of the distance embedding. The resulting messages are then summed for each node and normalized by the estimated average degree. The average degree depends on $r_{max}$, the cutoff distance for edges to be considered, and is estimated by interpolating between a set of known degrees for different values of $r_{max}$. The average degree is raised to the power $p = 0.7$ to better model the correlation between messages. Then the new node features are made by adding a linear combination of the old node features to the normalized messages. $r_{max}$ and the frequencies of the individual Bessel basis functions are trained as any other parameter.}
\label{fig:interaction}
\end{figure}
The interaction block, depicted in Figure \ref{fig:interaction}, elaborates on the basic convolution described in NequIP \cite{batzner2022nequip}. Node features and geometric information are combined using a Clebsch-Gordan tensor product, which fully characterizes all equivariant linear functions. Each output is weighted according to a filter generated from the distance associated with each edge, so the contribution of a specific component of the message varies smoothly with radius. Given the average number of neighbors $k$, we divide the sum of incoming messages by $k^{0.7}$, a middle ground between values of $\sqrt{k}$ and $k$ used by NequIP and MACE respectively. Instead of maintaining a single $k$, we compute this value for several choices of $r_{max}$ and then linearly interpolate between them to estimate the effective $k$ for any $r_{max}$. This lets us vary $r_{max}$ during training without adverse effects on normalization of messages, increasing the model's flexibility with only a single additional parameter.

Of particular note is the use of a simple linear layer instead of an MLP for the filter generation. We find that a simple linear layer has negligible loss of expressivity (99\% of variance explained for SevenNet-0's model checkpoint) and has dramatic performance benefits: the total compute, peak memory use, and parameter counts are all significantly improved. Further discussion of these design choices and their impact can be found in Methods.

\subsection*{Node self-interaction: the $S^2$-MLP-Mixer}
\begin{figure}[th!]
\includegraphics[width=\textwidth]{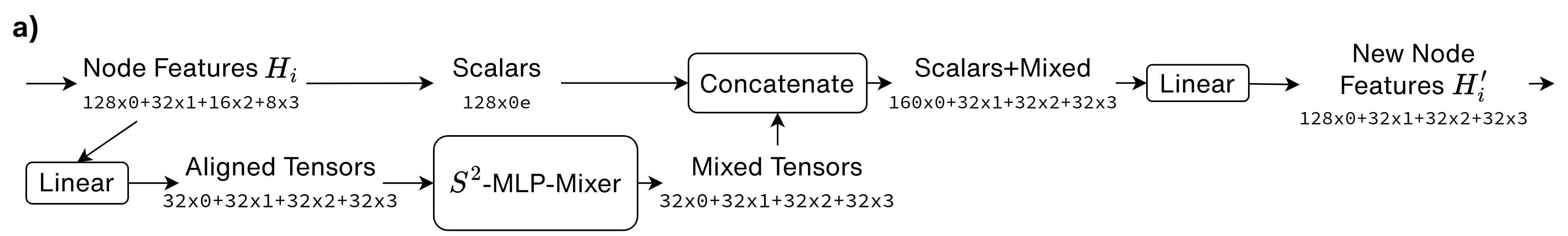}
\vspace*{0.5em}
\includegraphics[width=\textwidth]{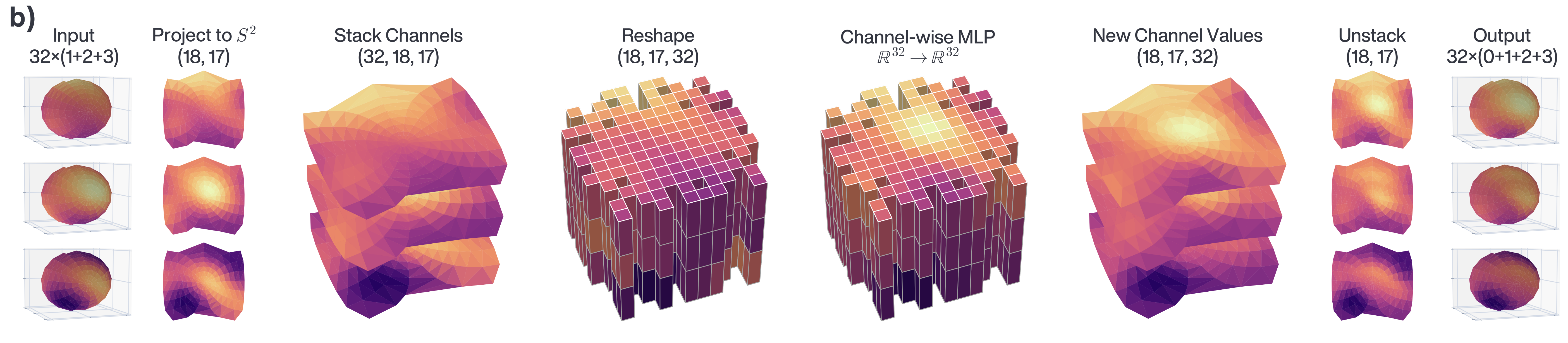}
\caption{a) The Facet self-interaction block. Following EquiformerV2, scalars are treated separately. The initial linear layer mixes information across channels of the same type and maps so the non-scalar inputs have the same count. To reduce overfitting and speed up the model, the input to the $S^2$-MLP-Mixer block is split into halves consisting of 16 groups of irreps each, processed separately, and then recombined. b) The Facet nonlinearity used in self-interaction, dubbed $S^2$-MLP-Mixer. Different irreps are combined and projected on the sphere $S^2$. This is then reinterpreted as a $N_A \times N_B \times C$ image, with $N_A, N_B = 18, 17$ controlling the fineness of the spherical grid. A channel-wise MLP in this space mixes inputs equivariantly, after which the outputs are projected back into the original basis.}
\label{fig:s2-mlp-visual}
\end{figure}

To process the information received during message passing and allow messages sent to the same node to mix, a self-interaction step is required: a node-wise update function that mixes information within a single environment. Standard GNNs \cite{choudhary2021alignn} apply an MLP. This works when the representation is entirely scalar, and so Facet uses a standard 2-layer MLP for the last layer's self-interaction, but standard MLPs are not equivariant and as such cannot process geometric information. Instead, Facet uses the correspondence between irreps and spherical harmonics, a basis of functions on the sphere $S^2$ (Figure \ref{fig:s2-mlp-visual} b). By projecting each input irrep to a function then discretized to a spherical grid, a collection of irreps can be interpreted as a spherical image with individual channels. Any function applied across the entire sphere corresponds to an equivariant function, treating all rotations equally. As such, a simple MLP can be applied channel-wise to mix the different irreps, after which the output is projected back into the space of the input.

The core idea of projection to $S^2$ is not new: it was proposed in \cite{cohen2018sphericalcnn} and incorporated into a property prediction model in EquiformerV2 \cite{liao2023equiformerv2} as $S^2$ activation. To indicate the use of a full MLP instead of a single nonlinearity applied to the channels, and to connect it with MLP-Mixer architectures used in the computer vision literature, we call our block $S^2$-MLP-Mixer. To our knowledge, it has yet to be used as the nonlinearity within a message-passing neural network, and we believe this choice offers several advantages over other common approaches. The MACE architecture uses a symmetric tensor-product layer, computing a linear function of $H_i + (H_i \otimes H_i) + (H_i \otimes H_i \otimes H_i) + \dots$ to a given max degree \cite{batatia2022designspace}. This operation is flexible, and can approximate any equivariant function through a Taylor approximation, but computing these tensor products is extremely expensive. NequIP and subsequent similar networks (GNoMe, SevenNet) use a nonlinear gate function, scaling each non-scalar by a weight derived from an MLP applied to the scalars. This operation is significantly cheaper, but only considers the norm of each non-scalar and does not mix non-scalars with one another. The $S^2$-MLP-Mixer block is nonlinear and mixes all dimensions of the input without using a full tensor product.

The computational difficulty in applying the $S^2$-MLP-Mixer is the need to apply the MLP across each of the $18 \times 17$ spherical grid locations used in Facet. We find a simple way to reduce the expense of the computation: because individual channels can be easily mixed with simple linear layers, we can split the input into heads and only mix information within a single head. 

Figure \ref{fig:s2-mlp-visual} a) shows how this nonlinearity is incorporated into the larger self-interaction block. Following EquiformerV2, the scalars are processed separately and recombined after mixing. Additional details of our Facet model are described in the Methods Section including representation, message generation, MLP convolution, message aggregation, node updating, element embeddings, and model readouts. 

\subsection*{Model Performance}
To fairly assess our design, we compare our models to other steerable GNNs trained on the MPTrj dataset containing 1,580,395 structures from the trajectories of 145,923 unique stable crystals.  

For objective evaluation, it should be noted that the two SevenNet models' test set performance may be over-estimated due to possible data leakage. SevenNet-0's performance is reported directly from the authors, specifically from the July 2024 checkpoint. They did not hold out any data, so it is difficult to estimate performance on new data. 

To more fairly compare SevenNet-0 to our work, we fine-tune starting from the SevenNet-0 checkpoint after making several \textit{post hoc} changes. We fit a spline to the convolution filter MLPs used in SevenNet, collapse the two output linear layers into a single one, and re-normalize the weights to match the average number of neighbors of our dataset. We set $r_{max}$ to the original value of 5 angstroms but, unlike SevenNet-0, let $r_{max}$ vary. The resulting model, SevenNet-Streamlined, is fine-tuned for 12 epochs on the same dataset splits and objective as Facet (MAE of formation energy). Because the original model was trained without holding out data, it is unclear how much residual knowledge from the checkpoint remains in the fine-tuned and modified version.
In addition, MACE-MP-0 refers to the "small" version MACE model trained for 250 epochs, of which the last 50 emphasize energy prediction. This makes it a good comparison for our model trained on energy alone. MACE-MP-0 was trained for 250 epochs on a cluster of 80 NVIDIA A100 GPUs.

In contrast, Facet-Small is trained from scratch, on energy prediction alone, for 25 epochs. We train starting from 3 random seeds. We report both the average performance of the individual models (Facet-Small-Average) and the performance of an ensemble averaging the predictions of the three models (Facet-Small-Ensemble).

\begin{figure}[h!]
\includegraphics[width=\textwidth]{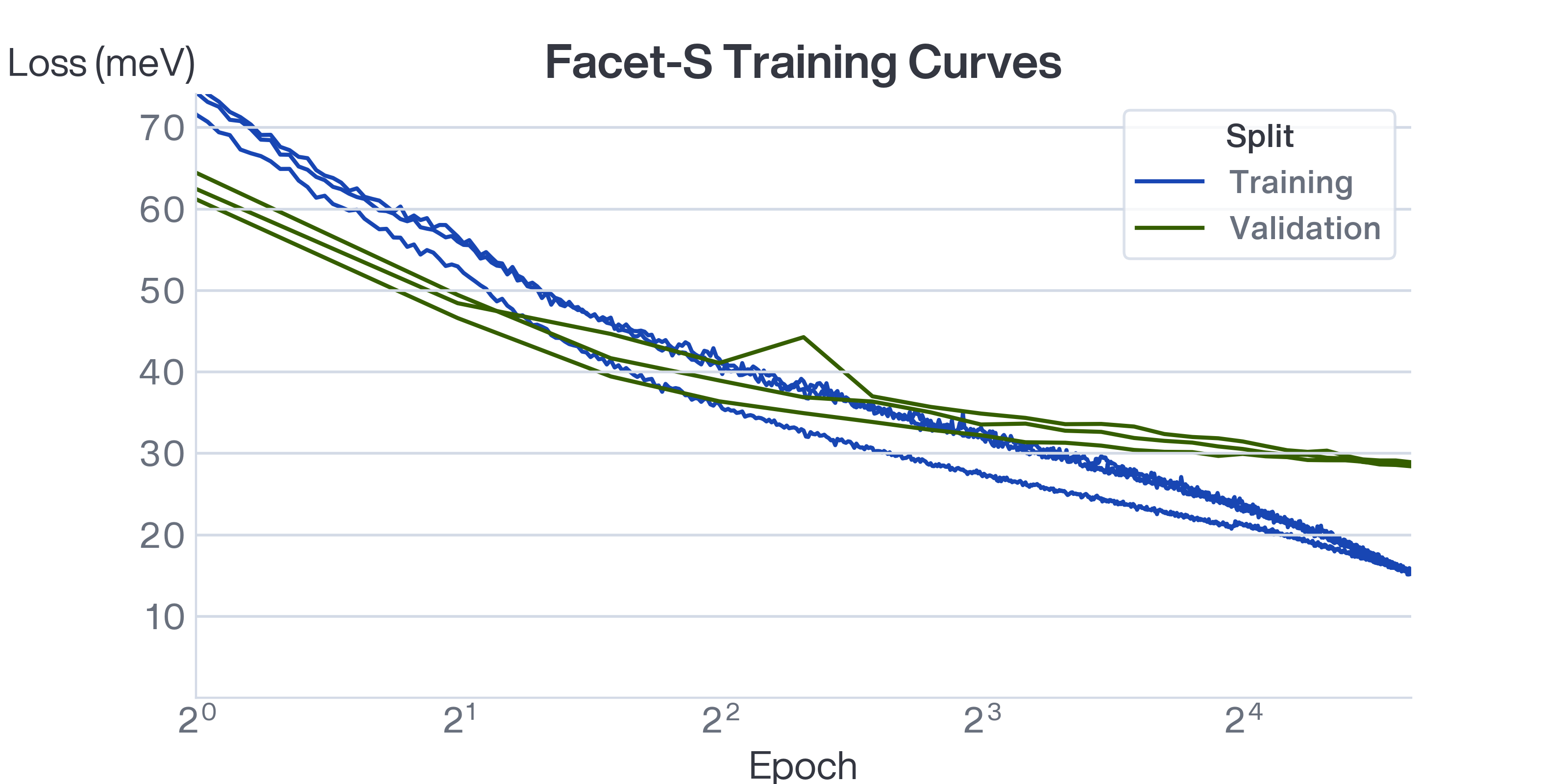}
\caption{Training curves for Facet-Small models, each starting from a different random initialization with all else equal and trained for 25 epochs. (Loss starts at approximately 470 meV: epoch 0 is omitted so training progress is more clearly visible). With the use of residual layers, normalization, and element-wise scaling, Facet-Small training is reliable and consistent.}
\label{fig:training-curve}
\end{figure}

For all models we train (including SevenNet-streamlined, Facet-Small-Average, and Facet-Small-Ensemble), we use EMA \cite{ema} with a decay of 0.99, applied every 32 steps, and a batch size of 32 structures. We use the Prodigy \cite{mishchenko2024prodigy} optimizer, which adaptively modifies the learning rate. The dataset is split 10:1:1 training:validation:test, such that structures from the same trajectory are put into the same split. This prevents leakage between the three sets. The training loss curve for each of the three Facet-Small models, shown in Figure \ref{fig:training-curve}, indicates training is relatively stable. 

The final performance comparison results are shown in Table \ref{tab:mptrj-results}. The top half shows the performance and training costs for selected baseline algorithms for structure-based crystal energy prediction. All models are trained with the same training set. We can observe the impact of large foundation models on improving prediction performance: EquiformerV2-S has the largest number of model parameters (~31M) compared to CHGNet's 412K parameters, while the former achieves the best performance (6 meV training error and 12.4 meV validation error, which are more than two times better than CHGNet's 26 meV and 29 meV, respectively). On the other hand, the MACE-MP-0 model has just about 12.3\% of EquiformerV2-S's model parameters while achieving similar validation error (13 meV compared to 12.4 meV), demonstrating its model efficiency. The SevenNet-0 model is even more efficient with 2.7\% of EquiformerV2-S's parameters while achieving 11 meV training error. While the top three models all achieved impressive prediction performance, their computational costs are significant, requiring from 90 days to 310 days of training on a single Nvidia A100 GPU. This makes it extremely challenging to optimize the hyperparameters for the models or to iterate model development. In addition, the large number of model parameters in EquiformerV2-S and MACE-MP-0 usually leads to slower inference, critical for long-range molecular dynamics simulations. Here, SevenNet-0 achieves the best balance between prediction accuracy and inference time, making it a good baseline for performance comparison with our Facet models.

\definecolor{ourmodelgray}{gray}{0.93}

\sisetup{
  table-format = 2.1, %
  table-space-text-post = *, %
}

\begin{table}[h!]
\caption{Comparison of formation energy prediction performance of Facet against baselines on MPTrj (meV). MACE-MP-0 refers to the "L0-energy" checkpoint trained in January 2024: results are from the authors' dataset splits. SevenNet-0 is reported directly from the authors (July 2024 checkpoint), who did not hold out test data.
*SevenNet-Streamlined trains with the same objective as Facet-S on the same data split, but some data leakage from the original checkpoint likely persists. Three Facet-Small models were trained: the average MAE and the performance of the ensemble are both reported. All computing costs are mapped to a single GPU.}
\begin{tabular}{l @{\hspace{1em}} r @{\hspace{1em}} l @{\hspace{1em}} S S S}
\toprule
{Model} & {Params} & {Training Compute} & {Train} & {Valid} & {Test} \\
\midrule

EquiformerV2-S \cite{barrosoluque2024omat} & 31,207,434 & 180-200 days w/A100 (estimated) &6 &12.4 & {N/A} \\
MACE-MP-0 & 3,847,696 & 310 days w/ A100 & 12.5 & 13.0 & {N/A} \\
SevenNet-0 (July 2024) & 842,748 & 90 days w/ A100 & 11 & {N/A} & {N/A} \\
CHGNet v0.3 \cite{zhou2024fastchgnet} & 412,525 & 8.3 days w/ A100 & 26 & 29 & 29 \\

\midrule

\rowcolor{ourmodelgray}
SevenNet-Streamlined & 616,253 & 90 days w/ A100 + 1 day w/ RTX 3090 & 10.9 & 20.7* & \textbf{20.1*} \\

\rowcolor{ourmodelgray}
Facet-Small (Average) & \textbf{270,383} & \textbf{2 days w/ RTX 3090} & 12.8 & 27.5 & 28.6 \\

\rowcolor{ourmodelgray}
Facet-Small (Ensemble) & 811,149 & 6 days w/ RTX 3090 & 10.8 & 24.6 & 23.1 \\

\bottomrule
\end{tabular}

\label{tab:mptrj-results}
\end{table}

The second half of Table 1 shows the performance of our Facet models compared to the closest baseline model SevenNet with the same validation and test sets. We find that our Facet-Small (ensemble) model achieves similar training error (10.8 meV versus 10.9 meV) with a similar number of model parameters, but was trained using only 6 days of RTX 3090 running time compared to the 90 days of A100 plus 1 day of RTX 3090 for SevenNet-streamlined. Its test error of 23.1 meV is also close to SevenNet-Streamlined's 20.7 meV (which may be overestimated due to data leakage in its pretraining stage). Our Facet-Small ensemble model achieves comparable formation energy prediction performance using only 1/15 of the training cost, a 93.3\% reduction in training time. In addition, we find that the Facet-Small average model also achieves strong performance with a test error of only 28.6 meV, which can be trained in only 2 days on an RTX 3090, demonstrating the high efficiency of our Facet model in both the number of model parameters and associated training cost (a 97.8\% reduction in training cost). This is similar to what DeepSeek-R1 has achieved in training a high-performance reasoning LLM. We believe the techniques we proposed in building the Facet model have great potential to improve other SOTA foundation models for atomic modeling.

\subsection*{Facet as a Fast Screening Potential for Crystal Structure Prediction}
One application of energy prediction models is as an objective function for crystal structure prediction (CSP) algorithms such as MAGUS \cite{wang2023magus}. Facet's accelerated training works well for procedures that use DFT as a reference and fine-tune a deep learning potential to screen candidate structures for DFT calculations.

In procedures without DFT, the potential becomes the bottleneck for crystal structure prediction: a more comprehensive search for an energy minimum demands faster inference. To evaluate Facet's performance in this scenario, we conducted a scaling test of various deep learning potentials for crystal structure prediction. We take three MACE-MP models written in PyTorch using the authors' library, the SevenNet-0 model implemented in the same JAX codebase as Facet-Small, and Facet-Small itself.

We consider two chemical systems: Ca-O-P and Fe-K-O. We take all of the stable structures in these systems from the Materials Project and consider both the original structure and a 2×2×2 supercell. This produces a range of atomic configurations at different scales indicative of a typical crystal structure prediction workflow.

\begin{figure}[h!]
\includegraphics[width=\textwidth]{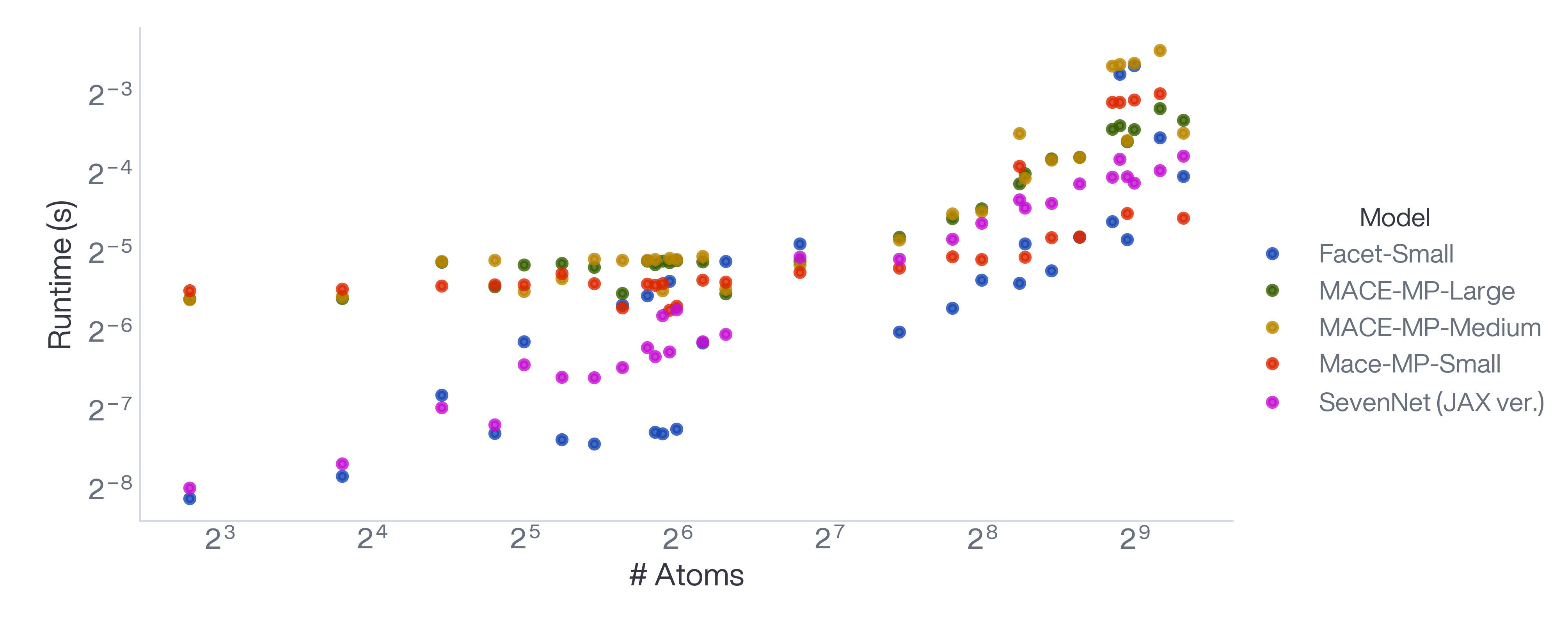}
\caption{Speed of relaxation for various crystal structures at differing scales. Especially for smaller structures, Facet-Small offers very low overhead, scaling well as size increases.}
\label{fig:scaling}
\end{figure}

We rattle the atomic coordinates randomly for each configuration and conduct relaxation for 100 steps for each structure, measuring the time each step requires. We plot the median time per step for each potential and input configuration (Figure \ref{fig:scaling}.) 

The effect of JAX can be clearly seen for small models: its compilation reduces overhead, which is quite significant for small configurations. As atomic configurations increase in scale, there are notable outliers—persistent across different runs of the experiment—that we attribute to missed low-level compilation optimizations and differences in the sparsity of atomic configurations. Despite this noise, the scaling relationships are still easily visible: as the cost of convolutions comes to dominate the runtime of the model, the differences between the models decrease (each model uses broadly the same convolution operation), although Facet-Small still remains efficient. 

Across all shown configurations, Facet has a significant performance edge. The mean performance edge across each configuration is a factor of 1.35 for SevenNet and 2.35/2.93/2.73 for MACE-MP Small/Medium/Large. While not as large as the orders-of-magnitude improvements in training time, the efficiency of Facet's architecture also accelerates inference for crystal structure prediction.

\subsection*{Facet Learns Informative Element Embeddings From Scratch}

Another way to understand the Facet model's performance is analyzing its learned elemental embeddings. 
Facet trains 128-dimensional embeddings for each element, initialized with no prior information, and uses them as the first node values before message passing begins.

We find that learning embeddings from scratch outperforms pretrained embeddings for final model performance. Specifically, we train a version of the Facet-Small model, substituting the learned embeddings for pre-trained crystal transformer universal atomic embeddings (ct-UAE) from \cite{jin2025ctuae} and leaving all else identical. Compared to our three Facet-Small models, which achieved an average 12.8/27.5/28.6 meV on train/valid/test data respectively, the version using ct-UAE has 14.6/30.5/31.9 meV, moderately worse across the board. In \cite{jin2025ctuae}, the authors find that MACE-MP similarly does not improve with their embeddings, but other models such as ALIGNN and CHGNet do exhibit improvements. We hypothesize that a sufficiently powerful backbone network obviates the need for pre-trained embeddings: our model seems able to learn the character of each element even more effectively than ct-UAE's model. A visualization using UMAP \cite{mcinnes2020umap} of Facet's learned embeddings in Supplementary File Figure S5 demonstrates Facet readily learns the structure of the periodic table without the need for explicit feature engineering.

\subsection*{MLP Convolutional Filters Are Simple Splines in Disguise}
The message filter, weighting each element by a scalar dependent on the distance between the sender and receiver, is the only mechanism by which distance information is included in the model network. It is also applied per edge, so achieving the optimal mix of expressivity and efficiency is paramount. Distance information is crucial for understanding the chemical environment, but redundant computation has a significant impact on performance within the filter.

Prior work (NequIP, GNoME, SevenNet, MACE) parameterizes $W$ as a multi-layer perceptron (MLP) mapping from a set of radial basis functions, with no biases and activation functions that map the origin to itself. This ensures that $W(0) = \vec{0}$ as required for the cutoff function to apply. We deviate from this multi-layer convention, which we show can significantly hamper network efficiency.

\begin{figure}[th!]
\includegraphics[width=\textwidth]{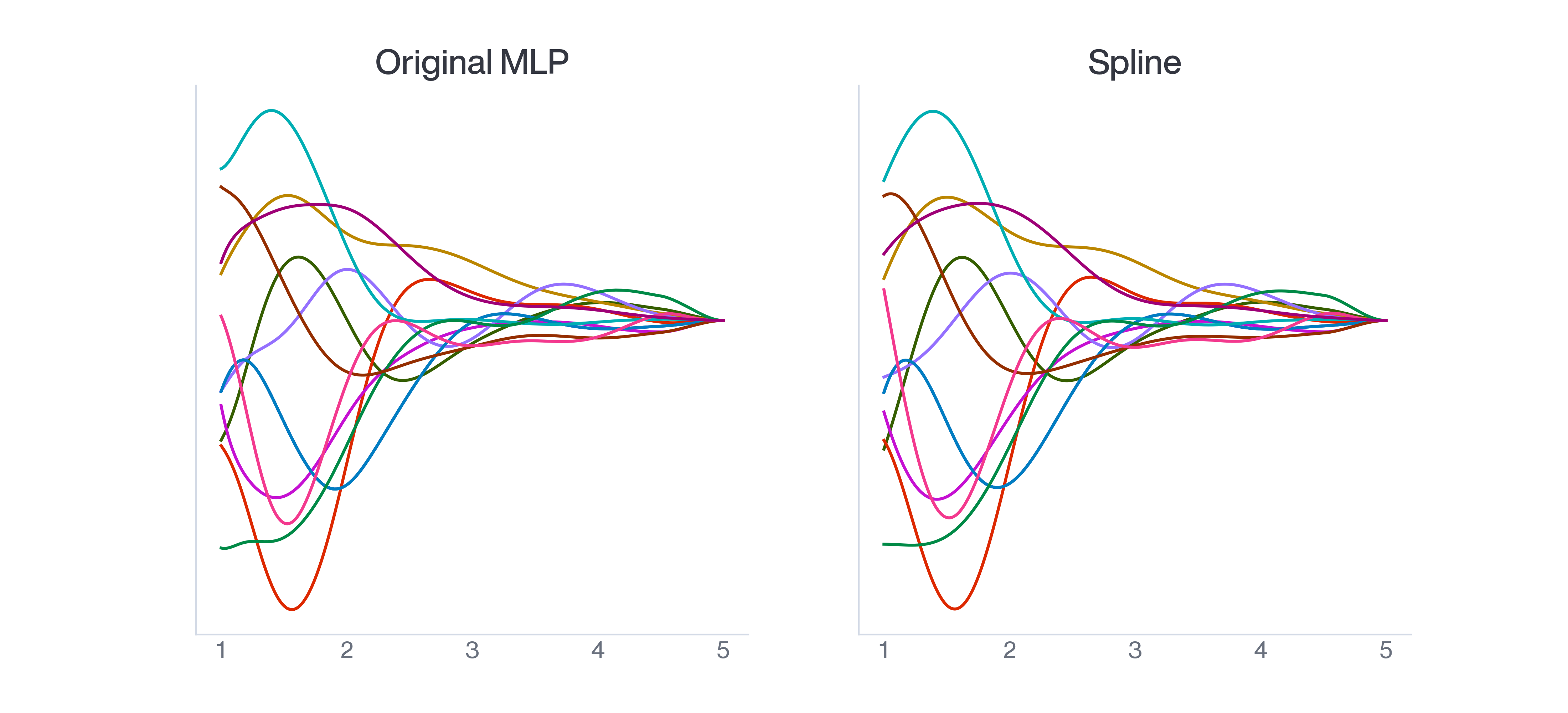}
\caption{Comparison of expensive MLP weight functions randomly selected from SevenNet-0 (left) and a simple linear combination of 8 Bessel basis functions fit to each curve (right), after applying SevenNet-0's distance envelope. Even a model trained using an MLP essentially chooses to use the simpler linear subspace representable by a spline.}
\label{fig:basis-comparison}
\end{figure}

No matter how complex our filter function $W$, at heart it maps a positive real to a vector of reals. Each output component can be considered independently. Consider a single channel of the output $W_i$. We expect $W_i$ to be broadly smooth: a small change to the input, the edge distance, should have a limited effect on the output.
A natural choice is then to consider $W$ as a spline. This corresponds to an MLP with no hidden layers or activation functions.
For SevenNet-Streamlined, we use the fact that we can easily enumerate the input space of the filter to fit a spline to the existing filters learned by SevenNet-0 as a 3-layer MLP. We fit a spline with 8 Bessel basis functions using linear regression. We achieve $R^2 > 0.99$, a nearly perfect fit (Figure \ref{fig:basis-comparison}). This does introduce a noticeable error in the final outputs, but with fine-tuning the model can adjust to the differences.

Our Facet architecture uses the same kind of filter: an 8-basis spline. We find that, not only does a spline achieve comparable results as shown in SevenNet-Streamlined, but the smoothness that it enforces provides a useful inductive bias during training, preventing overfitting on small differences in distance between atoms. Because filters can be refit quickly with orthogonal basis functions, future work could explore gradually increasing the resolution of the filter during training—tuning small differences in distance between atoms only as the network grows more accurate.
In both models, we let the frequencies of the basis functions train. This lets the model adjust the filter's fidelity in different areas of the input distribution with only eight additional parameters.

\section*{Discussion}
As the importance of MLIPs for materials discovery has become clear, increasingly large datasets have been developed, with increasingly large models to match \cite{liao2023equiformerv2,qu2024escaip,merchant2023gnome}. As dataset size becomes less of a bottleneck, efficient training of large potential models with limited computational resources becomes vital. Here we find that model architecture differences and optimizations can reduce the compute power needed to train and run inference on models by an order of magnitude without sacrificing performance. For screening large amounts of hypothetical materials or fine-tuning on new data, this speed directly accelerates downstream materials discovery applications.
There remain many avenues for future exploration. Scaling up the Facet architecture to larger models, datasets, and different training objectives should result in higher-accuracy potentials. The complex nonlinear behavior of the $S^2$-MLP-Mixer block complicates model interpretability: future analysis could shed new light on what makes the block effective and lead to improvements. Facet uses a simple information flow, optimizing mainly at the level of individual blocks of a message-passing network. More carefully incorporating node species, hypergraphs, or long-range interactions could address known issues of message-passing GNNs such as over-smoothing \cite{omee2022scalable,cai2020note} and further strengthen model performance.

In conclusion, through careful pruning of redundant computation and the application of gridded MLPs as general equivariant nonlinearities, we have developed the Facet architecture for steerable GNNs widely used in SOTA ML potentials. Through training a small model that compares favorably with much larger potentials in the prediction of energy, we demonstrate that property prediction models optimized for training and inference speed could continue to accelerate materials discovery and make the fruits of recent deep learning efforts more accessible to practitioners.

\section*{Methods}

\subsection*{Representations of $SO(3)$}
Before examining the Facet model design, we introduce compact notation following \texttt{e3nn} and review the principles that govern allowed operations on representations of $SO(3)$.

Different geometric quantities behave differently under rotation. Some quantities are invariant, such as distance: these are scalars, and they have degree 0. Other quantities behave as vectors, changing by a rotation matrix, and have degree 1. Higher-order degrees no longer correspond directly to conventional geometry: a nine-dimensional 3x3 tensor corresponding to \textit{e.g.,} stress can be decomposed into a degree-0 scalar (the trace), a degree-1 vector (the symmetric component), and a degree-2 tensor (the antisymmetric component). We focus here on the irreducible representations, abbreviated \textit{irreps}. A degree-$l$ irrep, or tensor, has $2l + 1$ dimensions, which we index by $-l \le m \le l$.

Consider, for example, two vectors $\langle a_1, b_1, c_1 \rangle$ and $\langle a_2, b_2, c_2 \rangle$. Many potential products exist that linearly combine the two vectors into a scalar. However, most are not equivariant. The only equivariant combination is, up to a constant scaling factor, the standard dot product: $a_1 a_2 + b_1 b_2 + c_1 c_2$.

The \textit{Clebsch-Gordan tensor product} generalizes this operation, describing all possible equivariant products of spherical tensors: the typical scalar and scalar-vector products, the dot product, the cross product, and so on for higher-degree tensors.
Not all combinations of tensors have equivariant combinations: this product is only nontrivial for input degrees $l_1, l_2$ and output degree $l_3$ when $|l_1 - l_2| \le l_3 \le l_1 + l_2$.
These tensors are representations of $SO(3)$, the group of proper rotations. Additionally modeling inversions simply requires tracking the parity of each tensor, odd or even, and then only allowing tensor products that follow the normal rules of parity. In this work, we use only even tensors.
A collection of tensors of different degrees is thus denoted using the number of tensors of a given degree, that degree, and \texttt{e} for even. A collection of 128 scalars, 64 vectors, and 32 degree-2 tensors is denoted \texttt{128x0e + 64x1e + 32x2e}.

There is a correspondence between irreps and an orthogonal basis of functions on the sphere $S^2$, the spherical harmonics. This correspondence allows us to project functions on the sphere onto irreps and vice versa. We use this to encode edge directions: the spherical harmonic representation of an edge is the embedding of a delta function with point mass aligned with the edge. The larger the order of tensors we use in that embedding, the more fidelity the model has to distinguish directional inputs within the network.

\subsection*{Message Generation}
In the ACE (Atomic Cluster Expansion) framework Facet uses, the network incorporates spatial information only through the representations of edge vectors and their effects on messages sent between nodes. Because the number of messages far exceeds the number of nodes, this part of the network needs to capture as much spatial information as possible efficiently.

Models such as EquiformerV2 \cite{liao2023equiformerv2} use attention as an alternate means of aggregating messages, but here we limit ourselves to typical GNN convolutions. NequIP \cite{batzner2022nequip}, GNoME \cite{merchant2023gnome}, and SevenNet \cite{park2024sevennet} use essentially identical messages, due originally to NequIP. 

For an edge with distance $r$ and spherical harmonics $Y$ between nodes with features $z_s$ and $z_t$ of the source and target node, in principle the message can be any function $M(r, Y, Z_s, Z_t)$. $Z_t$ is not considered: messages do not depend on the receiver.

To construct a message $M(r, Y, Z_s)$, the two sets of irreps are equivariantly combined via a tensor product (\ref{fig:conv-tp}). Each individual channel of the output is scaled by a weight controlling how information flows through the network. These weights are scalars computed from $r$. We denote this weighted tensor product $M(r, Y, Z_s) = Y \tens{W(r)} Z_s$. $W: \R^+ \to \R^n$, with $n$ the number of individual irreps in the output, is the only parameterized function, so we consider it in more detail.

Note that, if $W(r) = \vec{0}$, the message is also zero. This is how a differentiable distance cutoff is implemented. If we have a cutoff function $c: \R^+ \to \R$ such that $c(0) = 1$ and $c(r) = 0$ for all $r \ge r_{max}$, then we can use $W(r) = f(r c(r))$ for some function $f$ with $f(0) = \vec{0}$, and no edge beyond the cutoff will contribute to the output.

\subsection*{Message Aggregation}
In our Facet model, the messages coming into a node are summed and then normalized by a constant, as is standard in the literature: this prevents any scale issues during training. In MACE \cite{batatia2022designspace}, it was found that normalizing messages by the average number of neighbors $k$ instead of the square root $\sqrt{k}$ (as NequIP originally proposed) gives better performance. We shed light on this choice here.

Consider $k$ messages: because summation and normalization do not mix dimensions, it suffices to analyze the 1-dimensional case. Let us assume that the messages are Gaussian with mean 0 and standard deviation $\sigma$. The standard deviation of the sum of these messages depends on the correlation between them. In particular, the standard deviation of the sum of $k$ messages distributed as $N(0, \sigma^2)$ such that each message has correlation $\rho$ with each other message is $\sqrt{k \sigma^2 + k (k - 1) \rho \sigma^2}$. As $\rho$ varies from 0 (independence) to 1 (perfect correlation), the sum's scale ranges from $\sqrt{k} \sigma$ to $k \sigma$. 

Although this analysis makes oversimplifying assumptions, we find its intuition useful: the choice between $\sqrt{k}$ and $k$ represents a particular hypothesis about the correlation $\rho$ between messages to the same destination node, just as the choice of filter corresponds to a certain hypothesis about the maximum effect of small changes on distance between atoms.
SevenNet-0, which does not modify $r_{max}$ during training, lets us empirically estimate what the correlation between messages looks like for a trained model. Empirically, we find that in SevenNet-0 normalization by $k^p$ with $p \approx 0.7$ is a better fit to the data compared to $p = 0.5$ or $p = 1$, which both correspond to rather unrealistic expectations of message correlation: that messages to the same node are completely independent of each other, or that messages to the same node are perfectly correlated.

\subsection*{Training $r_{max}$, the maximum bond length}

We encode crystals as graphs using the 32 nearest neighbors for each atom. We then treat $r_{max}$, the cutoff radius, as a normal trainable parameter. This choice means our choice of normalization is not simply a difference in initialization but rather impacts the entire training run. In preprocessing, we compute the average number of neighbors for different values of $r_{max}$. We linearly interpolate between these values to estimate the average number of neighbors for any $r_{max}$ in a differentiable manner. Denoting this estimated average degree as $k(r_{max})$, We divide the sum of incoming messages by $k(r_{max})^{p}$, with $p = 0.7$ as a hyperparameter we set based on our SevenNet-0 analysis.

In SevenNet-Streamlined, we do the same, but to convert the original checkpoint we re-normalize the checkpoint with the normalization constant used in their data, starting at the same $r_{max} = 5$ as they do.

We find that, when $r_{max}$ is mutable, it tends to decrease sharply at the beginning of training, before the model has learned how to effectively incorporate edge information, and then slowly increases as training continues and the additional information can be smoothly incorporated. Future work could examine whether scheduling $r_{max}$ can achieve better results, similarly to how diffusion models change the mix of time steps used over training to adjust the difficulty of the task.

\subsection*{Node Update}
Between message-passing layers, we need to mix information within nodes by applying a node update $N(Z, e)$ with $Z$ the hidden state of a node and $e$ the node's element. 
NequIP and GNoME incorporate $e$ by applying a different function for each element with separate parameters: $N(Z, e) = N_e(Z)$. This results in a near-hundred-fold increase in parameters and the inability to share parameters across elements in a way that makes learning more difficult. Following SevenNet, we remove this indexing and instead rely on the initial node embeddings to incorporate elemental information: $N(Z, e) = N(Z)$. This makes the model smaller and more effective.

What remains is determining the network parameterizing $N$. This function should be nonlinear to add expressivity to the network, as it surrounds linear operations. MACE computes a polynomial of the input up to a small fixed degree: computing tensor products $Z, Z \otimes Z, Z \otimes Z \otimes Z$, etc., and then linearly combining into the desired size of the output. The connection between this product and many-order interactions, elucidated in \cite{batatia2022designspace}, provides a strong theoretical foundation for the layer. The cost of computing these tensor products scales poorly with the size of the network, however. 
SevenNet instead uses a gate activation: the norms of non-scalar elements are modified by a set of gate scalars derived from the scalar elements. This is significantly cheaper to compute, but does not mix non-scalars between each other, limiting the flow of directional information.
We use the $S^2$-MLP-Mixer layer described above in an attempt to find a middle ground. We elucidate some details here.

The spherical grid used is $18 \times 17$ with the quadrature described in \cite{kostelec2008soft}. EquiformerV2 uses $18 \times 18$ with a different quadrature. We find that, with this level of detail, the equivariance error (change in outputs from rotated versions of the inputs) is less than 1\%, of which much is due to the underlying tensor product implementations and not the $S^2$-MLP-Mixer layer.

\subsection*{Model Readout}
Sequencing message passing and node update layers produces rich internal representations for each node. For energy prediction, we represent energy as a sum of node energies and thus need to predict a single scalar from the node state.
Following SevenNet, we set the last message-passing layer to only produce scalars, effectively reducing the max degree already employed. A model head converts these scalars to a single output, summed across nodes.

In SevenNet-0, the readout MLP consists of two linear layers. This introduces redundancy: two multiplications by matrices of size $128 \times 64$ and $64 \times 1$ is equivalent to a single $128 \times 1$ matrix multiplication. Removing this redundancy further saves parameters and computation across the network with no change in model outputs. An alternative is to instead add a nonlinearity, producing a nonlinear MLP.

We empirically find that the complexity of the head has a large impact on model training and overfitting. SevenNet-0 demonstrates that models can be expressive without a nonlinear head, and more tightly constraining the output mitigates overfitting and ensures more consistent training. 

\section*{Implementation details}
Facet and SevenNet-Streamlined are implemented in JAX \cite{jax2018github}, using the \texttt{e3nn-jax} library \cite{e3nn} for working with irreps and Flax \cite{flax2020github} for neural network architecture.

\section*{Data availability}
The MPTrj dataset used in this work is available at \href{https://doi.org/10.6084/m9.figshare.23713842.v2}{https://doi.org/10.6084/m9.figshare.23713842.v2}. The code necessary to preprocess the data is available at \href{https://github.com/nicholas-miklaucic/facet}{https://github.com/nicholas-miklaucic/facet}.

\section*{Code availability}
All code necessary to replicate the results is available at \href{https://github.com/nicholas-miklaucic/facet}{https://github.com/nicholas-miklaucic/facet}.

\printbibliography

@article{wang2023magus,
    author = {Wang, Junjie and Gao, Hao and Han, Yu and Ding, Chi and Pan, Shuning and Wang, Yong and Jia, Qiuhan and Wang, Hui-Tian and Xing, Dingyu and Sun, Jian},
    title = {MAGUS: machine learning and graph theory assisted universal structure searcher},
    journal = {National Science Review},
    volume = {10},
    number = {7},
    pages = {nwad128},
    year = {2023},
    month = {05},
    issn = {2095-5138},
    doi = {10.1093/nsr/nwad128},
    url = {https://doi.org/10.1093/nsr/nwad128},
    eprint = {https://academic.oup.com/nsr/article-pdf/10/7/nwad128/50709989/nwad128.pdf},
}

@inproceedings{
xu2019wliso,
title={How Powerful are Graph Neural Networks?},
author={Keyulu Xu and Weihua Hu and Jure Leskovec and Stefanie Jegelka},
booktitle={International Conference on Learning Representations},
year={2019},
url={https://openreview.net/forum?id=ryGs6iA5Km},
}

@article{ema,
  title={Acceleration of stochastic approximation by averaging},
  author={Boris Polyak and Anatoli B. Juditsky},
  journal={Siam Journal on Control and Optimization},
  year={1992},
  volume={30},
  pages={838-855},
  url={https://api.semanticscholar.org/CorpusID:3548228}
}

@article{friederich2021machine,
  title={Machine-learned potentials for next-generation matter simulations},
  author={Friederich, Pascal and H{\"a}se, Florian and Proppe, Jonny and Aspuru-Guzik, Al{\'a}n},
  journal={Nature Materials},
  volume={20},
  number={6},
  pages={750--761},
  year={2021},
  publisher={Nature Publishing Group UK London}
}

@article{wang2024enhancing,
  title={Enhancing geometric representations for molecules with equivariant vector-scalar interactive message passing},
  author={Wang, Yusong and Wang, Tong and Li, Shaoning and He, Xinheng and Li, Mingyu and Wang, Zun and Zheng, Nanning and Shao, Bin and Liu, Tie-Yan},
  journal={Nature Communications},
  volume={15},
  number={1},
  pages={313},
  year={2024},
  publisher={Nature Publishing Group UK London}
}

@article{yang2025efficient,
  title={Efficient equivariant model for machine learning interatomic potentials},
  author={Yang, Ziduo and Wang, Xian and Li, Yifan and Lv, Qiujie and Chen, Calvin Yu-Chian and Shen, Lei},
  journal={npj Computational Materials},
  volume={11},
  number={1},
  pages={49},
  year={2025},
  publisher={Nature Publishing Group UK London}
}

@article{wang2024machine,
  title={Machine learning interatomic potential: Bridge the gap between small-scale models and realistic device-scale simulations},
  author={Wang, Guanjie and Wang, Changrui and Zhang, Xuanguang and Li, Zefeng and Zhou, Jian and Sun, Zhimei},
  journal={Iscience},
  year={2024},
  publisher={Elsevier}
}

@article{roberts2024machine,
  title={Machine learned interatomic potentials for ternary carbides trained on the AFLOW database},
  author={Roberts, Josiah and Rijal, Biswas and Divilov, Simon and Maria, Jon-Paul and Fahrenholtz, William G and Wolfe, Douglas E and Brenner, Donald W and Curtarolo, Stefano and Zurek, Eva},
  journal={npj Computational Materials},
  volume={10},
  number={1},
  pages={142},
  year={2024},
  publisher={Nature Publishing Group UK London}
}

@article{belli2025efficient,
  title={Efficient modelling of anharmonicity and quantum effects in PdCuH2 with machine learning potentials},
  author={Belli, Francesco and Zurek, Eva},
  journal={npj Computational Materials},
  volume={11},
  number={1},
  pages={87},
  year={2025},
  publisher={Nature Publishing Group UK London}
}

@article{omee2024crystal,
  title={Crystal structure prediction using neural network potential and age-fitness Pareto genetic algorithm},
  author={Omee, Sadman Sadeed and Wei, Lai and Hu, Ming and Hu, Jianjun},
  journal={Journal of Materials Informatics},
  year={2024},
  publisher={OAE Publishing Inc.}
}

@article{omee2025polymorphism,
  title={Polymorphism Crystal Structure Prediction with Adaptive Space Group Diversity Control},
  author={Omee, Sadman Sadeed and Wei, Lai and Dey, Sourin and Hu, Jianjun},
  journal={arXiv preprint arXiv:2506.11332},
  doi={10.48550/arXiv.2506.11332},
  year={2025}
}

@article{guo2025deepseek,
  title={Deepseek-r1: Incentivizing reasoning capability in llms via reinforcement learning},
  author={Guo, Daya and Yang, Dejian and Zhang, Haowei and Song, Junxiao and Zhang, Ruoyu and Xu, Runxin and Zhu, Qihao and Ma, Shirong and Wang, Peiyi and Bi, Xiao and others},
  journal={arXiv preprint arXiv:2501.12948},
  year={2025}
}

@article{deringer2019machine,
  title={Machine learning interatomic potentials as emerging tools for materials science},
  author={Deringer, Volker L and Caro, Miguel A and Cs{\'a}nyi, G{\'a}bor},
  journal={Advanced Materials},
  volume={31},
  number={46},
  pages={1902765},
  year={2019},
  publisher={Wiley Online Library}
}

@article{deng2025exploring,
  title={Exploring DeepSeek: A Survey on Advances, Applications, Challenges and Future Directions},
  author={Deng, Zehang and Ma, Wanlun and Han, Qing-Long and Zhou, Wei and Zhu, Xiaogang and Wen, Sheng and Xiang, Yang},
  journal={IEEE/CAA Journal of Automatica Sinica},
  volume={12},
  number={5},
  pages={872--893},
  year={2025},
  publisher={IEEE}
}

@article{wang2025review,
  title={A review of DeepSeek models' key innovative techniques},
  author={Wang, Chengen and Kantarcioglu, Murat},
  journal={arXiv preprint arXiv:2503.11486},
  year={2025}
}

@article{omee2022scalable,
  title={Scalable deeper graph neural networks for high-performance materials property prediction},
  author={Omee, Sadman Sadeed and Louis, Steph-Yves and Fu, Nihang and Wei, Lai and Dey, Sourin and Dong, Rongzhi and Li, Qinyang and Hu, Jianjun},
  journal={Patterns},
  volume={3},
  number={5},
  year={2022},
  publisher={Elsevier}
}

@article{cai2020note,
  title={A note on over-smoothing for graph neural networks},
  author={Cai, Chen and Wang, Yusu},
  journal={arXiv preprint arXiv:2006.13318},
  year={2020}
}

@article{fuchs2025chemtrain,
  title={chemtrain-deploy: A parallel and scalable framework for machine learning potentials in million-atom MD simulations},
  author={Fuchs, Paul and Chen, Weilong and Thaler, Stephan and Zavadlav, Julija},
  journal={arXiv preprint arXiv:2506.04055},
  year={2025}
}

@article{han2025distmlip,
  title={DistMLIP: A Distributed Inference Platform for Machine Learning Interatomic Potentials},
  author={Han, Kevin and Deng, Bowen and Farimani, Amir Barati and Ceder, Gerbrand},
  journal={arXiv preprint arXiv:2506.02023},
  year={2025}
}

@article{fedik2022extending,
  title={Extending machine learning beyond interatomic potentials for predicting molecular properties},
  author={Fedik, Nikita and Zubatyuk, Roman and Kulichenko, Maksim and Lubbers, Nicholas and Smith, Justin S and Nebgen, Benjamin and Messerly, Richard and Li, Ying Wai and Boldyrev, Alexander I and Barros, Kipton and others},
  journal={Nature Reviews Chemistry},
  volume={6},
  number={9},
  pages={653--672},
  year={2022},
  publisher={Nature Publishing Group UK London}
}

@article{han2025benchmarking,
  title={Benchmarking Universal Machine Learning Interatomic Potentials for Real-Time Analysis of Inelastic Neutron Scattering Data},
  author={Han, Bowen and Cheng, Yongqiang},
  journal={arXiv preprint arXiv:2506.01860},
  year={2025}
}

@article{riebesell2023matbench,
  title={Matbench Discovery--A framework to evaluate machine learning crystal stability predictions},
  author={Riebesell, Janosh and Goodall, Rhys EA and Benner, Philipp and Chiang, Yuan and Deng, Bowen and Lee, Alpha A and Jain, Anubhav and Persson, Kristin A},
  journal={arXiv preprint arXiv:2308.14920},
  year={2023}
}

@article{chen2022universal,
  title={A universal graph deep learning interatomic potential for the periodic table},
  author={Chen, Chi and Ong, Shyue Ping},
  journal={Nature Computational Science},
  volume={2},
  number={11},
  pages={718--728},
  year={2022},
  publisher={Nature Publishing Group US New York}
}

@article{fu2025learning,
  title={Learning smooth and expressive interatomic potentials for physical property prediction},
  author={Fu, Xiang and Wood, Brandon M and Barroso-Luque, Luis and Levine, Daniel S and Gao, Meng and Dzamba, Misko and Zitnick, C Lawrence},
  journal={arXiv preprint arXiv:2502.12147},
  year={2025}
}

@article{wen2022deep,
  title={Deep potentials for materials science},
  author={Wen, Tongqi and Zhang, Linfeng and Wang, Han and Srolovitz, David J and others},
  journal={Materials Futures},
  volume={1},
  number={2},
  pages={022601},
  year={2022},
  publisher={IOP Publishing}
}

@article{ramprasad2017machine,
  title={Machine learning in materials informatics: recent applications and prospects},
  author={Ramprasad, Rampi and Batra, Rohit and Pilania, Ghanshyam and Mannodi-Kanakkithodi, Arun and Kim, Chiho},
  journal={npj Computational Materials},
  volume={3},
  number={1},
  pages={54},
  year={2017},
  publisher={Nature Publishing Group UK London}
}

@article{louie2021discovering,
  title={Discovering and understanding materials through computation},
  author={Louie, Steven G and Chan, Yang-Hao and da Jornada, Felipe H and Li, Zhenglu and Qiu, Diana Y},
  journal={Nature Materials},
  volume={20},
  number={6},
  pages={728--735},
  year={2021},
  publisher={Nature Publishing Group UK London}
}

@article{neugebauer2013density,
  title={Density functional theory in materials science},
  author={Neugebauer, J{\"o}rg and Hickel, Tilmann},
  journal={Wiley Interdisciplinary Reviews: Computational Molecular Science},
  volume={3},
  number={5},
  pages={438--448},
  year={2013},
  publisher={Wiley Online Library}
}

@article{oganov2019structure,
  title={Structure prediction drives materials discovery},
  author={Oganov, Artem R and Pickard, Chris J and Zhu, Qiang and Needs, Richard J},
  journal={Nature Reviews Materials},
  volume={4},
  number={5},
  pages={331--348},
  year={2019},
  publisher={Nature Publishing Group UK London}
}

@misc{barrosoluque2024omat,
  title = {Open {{Materials}} 2024 ({{OMat24}}) {{Inorganic Materials Dataset}} and {{Models}}},
  author = {{Barroso-Luque}, Luis and Shuaibi, Muhammed and Fu, Xiang and Wood, Brandon M. and Dzamba, Misko and Gao, Meng and Rizvi, Ammar and Zitnick, C. Lawrence and Ulissi, Zachary W.},
  year = {2024},
  month = oct,
  number = {arXiv:2410.12771},
  eprint = {2410.12771},
  publisher = {arXiv},
  doi = {10.48550/arXiv.2410.12771},
  urldate = {2025-03-19},
  archiveprefix = {arXiv}
}

@misc{batatia2022designspace,
  title = {The {{Design Space}} of {{E}}(3)-{{Equivariant Atom-Centered Interatomic Potentials}}},
  author = {Batatia, Ilyes and Batzner, Simon and Kov{\'a}cs, D{\'a}vid P{\'e}ter and Musaelian, Albert and Simm, Gregor N. C. and Drautz, Ralf and Ortner, Christoph and Kozinsky, Boris and Cs{\'a}nyi, G{\'a}bor},
  year = {2022},
  month = nov,
  number = {arXiv:2205.06643},
  eprint = {2205.06643},
  publisher = {arXiv},
  doi = {10.48550/arXiv.2205.06643},
  urldate = {2024-10-30},
  archiveprefix = {arXiv}
}

@misc{batatia2024macemp0,
  title = {A Foundation Model for Atomistic Materials Chemistry},
  author = {Batatia, Ilyes and Benner, Philipp and Chiang, Yuan and Elena, Alin M. and Kov{\'a}cs, D{\'a}vid P. and Riebesell, Janosh and Advincula, Xavier R. and Asta, Mark and Avaylon, Matthew and Baldwin, William J. and Berger, Fabian and Bernstein, Noam and Bhowmik, Arghya and Blau, Samuel M. and C{\u a}rare, Vlad and Darby, James P. and De, Sandip and Pia, Flaviano Della and Deringer, Volker L. and Elijo{\v s}ius, Rokas and {El-Machachi}, Zakariya and Falcioni, Fabio and Fako, Edvin and Ferrari, Andrea C. and {Genreith-Schriever}, Annalena and George, Janine and Goodall, Rhys E. A. and Grey, Clare P. and Grigorev, Petr and Han, Shuang and Handley, Will and Heenen, Hendrik H. and Hermansson, Kersti and Holm, Christian and Jaafar, Jad and Hofmann, Stephan and Jakob, Konstantin S. and Jung, Hyunwook and Kapil, Venkat and Kaplan, Aaron D. and Karimitari, Nima and Kermode, James R. and Kroupa, Namu and Kullgren, Jolla and Kuner, Matthew C. and Kuryla, Domantas and Liepuoniute, Guoda and Margraf, Johannes T. and Magd{\u a}u, Ioan-Bogdan and Michaelides, Angelos and Moore, J. Harry and Naik, Aakash A. and Niblett, Samuel P. and Norwood, Sam Walton and O'Neill, Niamh and Ortner, Christoph and Persson, Kristin A. and Reuter, Karsten and Rosen, Andrew S. and Schaaf, Lars L. and Schran, Christoph and Shi, Benjamin X. and Sivonxay, Eric and Stenczel, Tam{\'a}s K. and Svahn, Viktor and Sutton, Christopher and Swinburne, Thomas D. and Tilly, Jules and van der Oord, Cas and {Varga-Umbrich}, Eszter and Vegge, Tejs and Vondr{\'a}k, Martin and Wang, Yangshuai and Witt, William C. and Zills, Fabian and Cs{\'a}nyi, G{\'a}bor},
  year = {2024},
  month = mar,
  number = {arXiv:2401.00096},
  eprint = {2401.00096},
  publisher = {arXiv},
  doi = {10.48550/arXiv.2401.00096},
  urldate = {2024-10-30},
  archiveprefix = {arXiv}
}

@article{batzner2022nequip,
  title = {E(3)-Equivariant Graph Neural Networks for Data-Efficient and Accurate Interatomic Potentials},
  author = {Batzner, Simon and Musaelian, Albert and Sun, Lixin and Geiger, Mario and Mailoa, Jonathan P. and Kornbluth, Mordechai and Molinari, Nicola and Smidt, Tess E. and Kozinsky, Boris},
  year = {2022},
  month = may,
  journal = {Nature Communications},
  volume = {13},
  number = {1},
  pages = {2453},
  issn = {2041-1723},
  doi = {10.1038/s41467-022-29939-5},
  urldate = {2024-11-04},
  langid = {english}
}

@article{choudhary2021alignn,
  title = {Atomistic {{Line Graph Neural Network}} for Improved Materials Property Predictions},
  author = {Choudhary, Kamal and DeCost, Brian},
  year = {2021},
  month = nov,
  journal = {npj Computational Materials},
  volume = {7},
  number = {1},
  pages = {185},
  issn = {2057-3960},
  doi = {10.1038/s41524-021-00650-1},
  urldate = {2025-01-20},
  langid = {english}
}

@inproceedings{cohen2018sphericalcnn,
  title = {Spherical {{CNNs}}},
  booktitle = {International {{Conference}} on {{Learning Representations}}},
  author = {Cohen, Taco S. and Geiger, Mario and K{\"o}hler, Jonas and Welling, Max},
  year = {2018},
  month = feb,
  urldate = {2025-01-26},
  langid = {english}
}

@misc{e3nn,
  title = {Euclidean Neural Networks: E3nn},
  author = {Geiger, Mario and Smidt, Tess and M., Alby and Miller, Benjamin Kurt and Boomsma, Wouter and Dice, Bradley and Lapchevskyi, Kostiantyn and Weiler, Maurice and Tyszkiewicz, Micha{\l} and Batzner, Simon and Madisetti, Dylan and Uhrin, Martin and Frellsen, Jes and Jung, Nuri and Sanborn, Sophia and Wen, Mingjian and Rackers, Josh and R{\o}d, Marcel and Bailey, Michael},
  year = {2022},
  month = apr,
  doi = {10.5281/zenodo.6459381},
  howpublished = {https://doi.org/10.5281/zenodo.6459381}
}

@misc{flax2020github,
  title = {Flax: {{A}} Neural Network Library and Ecosystem for {{JAX}}},
  author = {Heek, Jonathan and Levskaya, Anselm and Oliver, Avital and Ritter, Marvin and Rondepierre, Bertrand and Steiner, Andreas and {van Zee}, Marc},
  year = {2024},
  howpublished = {http://github.com/google/flax}
}

@inproceedings{gasteiger2020dimenetpp,
  title = {Fast and Uncertainty-Aware Directional Message Passing for Non-Equilibrium Molecules},
  booktitle = {Machine Learning for Molecules Workshop, {{NeurIPS}}},
  author = {Gasteiger, Johannes and Giri, Shankari and Margraf, Johannes T. and G{\"u}nnemann, Stephan},
  year = {2020}
}

@inproceedings{gasteiger2021gemnet,
  title = {{{GemNet}}: {{Universal Directional Graph Neural Networks}} for {{Molecules}}},
  shorttitle = {{{GemNet}}},
  booktitle = {Advances in {{Neural Information Processing Systems}}},
  author = {Gasteiger, Johannes and Becker, Florian and G{\"u}nnemann, Stephan},
  year = {2021},
  volume = {34},
  pages = {6790--6802},
  publisher = {Curran Associates, Inc.},
  urldate = {2025-01-20}
}

@misc{jax2018github,
  title = {{{JAX}}: Composable Transformations of {{Python}}+{{NumPy}} Programs},
  author = {Bradbury, James and Frostig, Roy and Hawkins, Peter and Johnson, Matthew James and Leary, Chris and Maclaurin, Dougal and Necula, George and Paszke, Adam and VanderPlas, Jake and {Wanderman-Milne}, Skye and Zhang, Qiao},
  year = {2018},
  howpublished = {http://github.com/jax-ml/jax}
}

@article{jin2025ctuae,
  title = {Transformer-Generated Atomic Embeddings to Enhance Prediction Accuracy of Crystal Properties with Machine Learning},
  author = {Jin, Luozhijie and Du, Zijian and Shu, Le and Cen, Yan and Xu, Yuanfeng and Mei, Yongfeng and Zhang, Hao},
  year = {2025},
  month = jan,
  journal = {Nature Communications},
  volume = {16},
  number = {1},
  pages = {1210},
  publisher = {Nature Publishing Group},
  issn = {2041-1723},
  doi = {10.1038/s41467-025-56481-x},
  urldate = {2025-03-21},
  copyright = {2025 The Author(s)},
  langid = {english}
}

@article{kostelec2008soft,
  title = {{{FFTs}} on the {{Rotation Group}}},
  author = {Kostelec, Peter J. and Rockmore, Daniel N.},
  year = {2008},
  month = apr,
  journal = {Journal of Fourier Analysis and Applications},
  volume = {14},
  number = {2},
  pages = {145--179},
  issn = {1531-5851},
  doi = {10.1007/s00041-008-9013-5},
  urldate = {2025-03-19},
  langid = {english}
}

@article{liao2022equiformer,
  title = {Equiformer: {{Equivariant Graph Attention Transformer}} for {{3D Atomistic Graphs}}},
  shorttitle = {Equiformer},
  author = {Liao, Yi-Lun and Smidt, Tess},
  year = {2022},
  publisher = {arXiv},
  doi = {10.48550/ARXIV.2206.11990},
  urldate = {2024-11-04},
  copyright = {Creative Commons Attribution 4.0 International}
}

@article{liao2023equiformerv2,
  title = {{{EquiformerV2}}: {{Improved Equivariant Transformer}} for {{Scaling}} to {{Higher-Degree Representations}}},
  shorttitle = {{{EquiformerV2}}},
  author = {Liao, Yi-Lun and Wood, Brandon and Das, Abhishek and Smidt, Tess},
  year = {2023},
  publisher = {arXiv},
  doi = {10.48550/ARXIV.2306.12059},
  urldate = {2024-11-04},
  copyright = {Creative Commons Attribution 4.0 International}
}

@misc{mcinnes2020umap,
  title = {{{UMAP}}: {{Uniform Manifold Approximation}} and {{Projection}} for {{Dimension Reduction}}},
  shorttitle = {{{UMAP}}},
  author = {McInnes, Leland and Healy, John and Melville, James},
  year = {2020},
  month = sep,
  number = {arXiv:1802.03426},
  eprint = {1802.03426},
  publisher = {arXiv},
  doi = {10.48550/arXiv.1802.03426},
  urldate = {2025-03-23},
  archiveprefix = {arXiv}
}

@article{merchant2023gnome,
  title = {Scaling Deep Learning for Materials Discovery},
  author = {Merchant, Amil and Batzner, Simon and Schoenholz, Samuel S. and Aykol, Muratahan and Cheon, Gowoon and Cubuk, Ekin Dogus},
  year = {2023},
  month = dec,
  journal = {Nature},
  volume = {624},
  number = {7990},
  pages = {80--85},
  publisher = {Nature Publishing Group},
  issn = {1476-4687},
  doi = {10.1038/s41586-023-06735-9},
  urldate = {2025-01-13},
  copyright = {2023 The Author(s)},
  langid = {english}
}

@inproceedings{mishchenko2024prodigy,
  title = {Prodigy: {{An}} Expeditiously Adaptive Parameter-Free Learner},
  booktitle = {Forty-First International Conference on Machine Learning},
  author = {Mishchenko, Konstantin and Defazio, Aaron},
  year = {2024}
}

@article{park2024sevennet,
  title = {Scalable {{Parallel Algorithm}} for {{Graph Neural Network Interatomic Potentials}} in {{Molecular Dynamics Simulations}}},
  author = {Park, Yutack and Kim, Jaesun and Hwang, Seungwoo and Han, Seungwu},
  year = {2024},
  month = jun,
  journal = {Journal of Chemical Theory and Computation},
  volume = {20},
  number = {11},
  pages = {4857--4868},
  publisher = {American Chemical Society},
  issn = {1549-9618},
  doi = {10.1021/acs.jctc.4c00190}
}

@inproceedings{qu2024escaip,
  title = {The {{Importance}} of {{Being Scalable}}: {{Improving}} the {{Speed}} and {{Accuracy}} of {{Neural Network Interatomic Potentials Across Chemical Domains}}},
  shorttitle = {The {{Importance}} of {{Being Scalable}}},
  author = {Qu, Eric and Krishnapriyan, Aditi S.},
  year = {2024},
  month = oct,
  urldate = {2024-11-04}
}

@article{ruff2024cogn,
  title = {Connectivity Optimized Nested Line Graph Networks for Crystal Structures},
  author = {Ruff, Robin and Reiser, Patrick and St{\"u}hmer, Jan and Friederich, Pascal},
  year = {2024},
  journal = {Digital Discovery},
  volume = {3},
  number = {3},
  pages = {594--601},
  publisher = {Royal Society of Chemistry},
  doi = {10.1039/D4DD00018H},
  urldate = {2024-10-30},
  langid = {english}
}

@article{xie2017cgcnn,
  title = {Crystal {{Graph Convolutional Neural Networks}} for an {{Accurate}} and {{Interpretable Prediction}} of {{Material Properties}}},
  author = {Xie, Tian and Grossman, Jeffrey C.},
  year = {2018},
  month = apr,
  journal = {Physical Review Letters},
  volume = {120},
  number = {14},
  pages = {145301},
  issn = {0031-9007, 1079-7114},
  doi = {10.1103/PhysRevLett.120.145301},
  urldate = {2025-01-20},
  langid = {english}
}

@article{yan2024periodic,
  title = {Periodic {{Graph Transformers}} for {{Crystal Material Property Prediction}}},
  author = {Yan, Keqiang and Liu, Yi and Lin, Yuchao and Ji, Shuiwang},
  year = {2022},
  month = dec,
  journal = {Advances in Neural Information Processing Systems},
  volume = {35},
  pages = {15066--15080},
  urldate = {2024-10-30},
  langid = {english}
}

@misc{zhou2024fastchgnet,
  title = {{{FastCHGNet}}: {{Training}} One {{Universal Interatomic Potential}} to 1.5 {{Hours}} with 32 {{GPUs}}},
  shorttitle = {{{FastCHGNet}}},
  author = {Zhou, Yuanchang and Hu, Siyu and Wang, Chen and Wang, Lin-Wang and Tan, Guangming and Jia, Weile},
  year = {2024},
  month = dec,
  number = {arXiv:2412.20796},
  eprint = {2412.20796},
  publisher = {arXiv},
  doi = {10.48550/arXiv.2412.20796},
  urldate = {2025-02-17},
  archiveprefix = {arXiv}
}

\section*{Acknowledgement}
The research reported in this work was supported in part by National Science Foundation under the grant and 2110033, 2311202, 2320292, and OAC-2311203.
The views, perspectives, and content do not necessarily represent the official views of the NSF. 

\section*{Author contributions}
J.H. initiated the project, N.M.conceived the Facet model, implemented the software and conducted all software experiments under the guidance of J.H. L.W., R.D., N.F., S.S.O., Q.L., S.D., and V.F. participated in the discussion of the Facet model and evaluation of the models. N.F. and R.D. helped to polish the table results. 
All authors contributed to writing and revising the manuscript.

\section*{Competing interests}
The authors declare no competing interests.

\end{document}